\documentclass[letterpaper,journal]{IEEEtran}
\usepackage{float}
\usepackage{algorithmic}
\usepackage{algorithm}
\usepackage{array}
\usepackage[caption=false,font=normalsize,labelfont=sf,textfont=sf]{subfig}
\usepackage{color}
\usepackage{textcomp}
\usepackage{stfloats}
\usepackage{url}
\usepackage{mathtools}
\usepackage{verbatim}
\usepackage{amssymb}
\usepackage{graphicx}
\usepackage{cite}
\usepackage{amsmath,amsfonts,amsthm}

\newtheorem{theorem}{Theorem}

\newtheorem{remark}{Remark}
\newtheorem{corollary}{Corollary}

\newcommand{\sphbess}{j}
\newcommand{\Ylm}{Y_{l}^q}

\newcommand{\PSI}[4]{\psi^{#1,#2}_{#3(#4)}}
\newcommand{\jj}{\mathrm{j}} 

\begin{document}

\title{Spherical Tessellation of Circular Arrays for Massive MIMO: Favourable Propagation Conditions and Spectral Design}

\author{Geili~T.~A.~El~Sanousi, Tuan~Anh~Le, Raed~A.~A.~Abd-Alhameed, Xin-She~Yang, and Huan~X.~Nguyen\vspace{-9mm}
\thanks{G.~T.~A.~El~Sanousi, T. A. Le, X.-S. Yang and H. X. Nguyen are with the Faculty of Science and Technology, Middlesex University, The Burroughs, Hendon, London, NW4 4BT, U. K. Email: \{ge174@live.mdx.ac.uk\},\{t.le; x.yang,h.nguyen\}@mdx.ac.uk. H.~X.~Nguyen is also with VinUniversity, Hanoi, 10000, Vietnam.}
\thanks{R.~A.~A.~Alhameed is with the University of  Bradford, U.K and Al-Farqadein University, Basrah 61004, Iraq. Email: r.a.a.abd@bradford.ac.uk.}
}

\markboth{IEEE WIRELESS COMMUNICATIONS LETTERS,~Accepted, 16 November~2026}{}

\maketitle
\bstctlcite{IEEEexample:BSTcontrol}
\begin{abstract}
Upper-mid-band XL-MIMO can exploit line-of-sight propagation, but periodic uniform cylindrical arrays may produce grating ridges. We analyse a spherically tessellated concentric circular antenna array with non-separable facet orientations. Under fixed normalised chord spacing and ideal facet visibility, we prove that user channels become asymptotically orthogonal and matched-filter SINR approaches the single-user SNR. Simulations confirm interference decay and SINR convergence. A spectral comparison shows that facet-normal diversity prevents the full-amplitude grating ridges of the cylindrical baseline, with the inter-facet residual decaying with array size. \vspace{-1mm}
\end{abstract}
\begin{IEEEkeywords}
Massive MIMO, favourable propagation, spherical tessellated concentric circular antenna arrays~(ST-CCAA).
\vspace{-4mm}
\end{IEEEkeywords}
\vspace{-5mm}
\section{Introduction}
\IEEEPARstart{R}{ecent} advances in extremely large-scale MIMO (XL-MIMO) and upper mid-band frequencies have shifted the paradigm of line-of-sight (LOS) channels from rank-deficient bottlenecks to high-capacity resources \cite{tian,bjorn}. Aperture expansion enables near-field operation, in which spherical-wavefront curvature allows beam focusing to multiplex users at identical angles but different depths \cite{cui,wymee,bjorn}. Conformal geometries such as uniform circular and cylindrical arrays (UCAs/UCLAs) exhibit favourable propagation (FP), meaning user channels become mutually orthogonal as the antenna count increases \cite{firoz,rivi}. While FP has been proven for UCAs and extended to cylindrical arrays via Kronecker decomposition \cite{firoz}, periodic vertical stacking in UCLAs can lead to coherent aliasing and grating ridges. However, extending FP to non-separable spherical geometries, where Kronecker decomposition is inapplicable, remains an open challenge. This letter analyses the Spherical Tessellated Concentric Circular Antenna Array (ST-CCAA), constructed by tessellating planar concentric circular antenna arrays (CCAA) \cite{dess, FHSA} along a spherical frame. Under the far-field LOS model in \cite{firoz}, we provide the orthogonality results necessary for subsequent near-field extensions. The main contributions are:

1)~We prove pointwise FP for the ST-CCAA under specific facet shielding and angle conditions. By employing a per-facet triangle inequality, the problem is reduced to single-ring UCA sub-problems, and Bessel bounding establishes SINR convergence to the single-user bound.

2)~We introduce the Poisson summation formula (PSF) applied to a spherical Bessel expansion of phase differences, revealing that, unlike the periodic UCLA lattice, the ST-CCAA facet assembly on the unit sphere \(S^2\) suppresses grating ridges.

3)~Through direct array-factor analysis, we attribute the absence of grating lobes to 3-D facet-normal diversity, which prevents the coherent aliasing alignment inherent in the periodic structure of UCLAs.

\emph{\textbf{Notation}:~$x, X$: scalars; $\mathbf{x}$: column vector; $\mathbf{X}$: matrix. $(\cdot)^*$, $(\cdot)^T$, $(\cdot)^H$, $\|\cdot\|$: conjugate, transpose, conjugate-transpose, and Euclidean norm. $|\cdot|$: absolute value or set cardinality. $\mathrm{vol}(\mathcal{L})$, $\lceil \cdot \rceil$, $\mathbb{E}[\cdot]$: volume of $\mathcal{L}$ (Lebesgue measure), ceiling operator, and expectation. $\mathbf{I}_x$: $x \times x$ identity matrix. $\mathcal{O}(\cdot)$, $\Omega(\cdot)$, and $\Theta(\cdot)$: asymptotic upper, lower, and tight bounds, i.e., $f{=}\mathcal{O}(g)$, $\Omega(g)$, and $\Theta(g)$ if $|f|\le c_2\,g$, $|f|\ge c_1\,g$, or $c_1\,g\le |f| \le c_2\,g$, respectively, for sufficiently large arguments, with $g>0$ and constants $c_1,c_2>0$.}
\vspace{-3mm}
\section{Channel Model}
\begin{figure}[ht]
    \centering
    \includegraphics[width=.3\textwidth,height=59 mm]{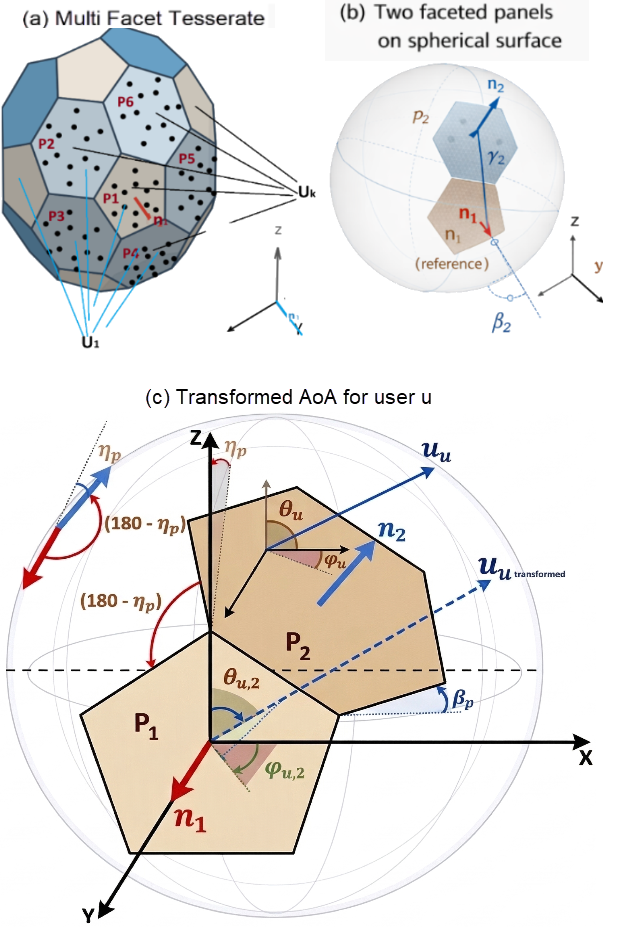}  \vspace{-2mm}
    \caption{(a)~Multiplanar ST-CCAA with \(P\) facets, each hosting a CCAA aperture. (b)~Reference facet (\(p=1\)), normal \(\mathbf{n}_1\), and facet (\(p=2\)), normal \(\mathbf{n}_2\), with tilt \(\eta_2\) and azimuth \(\beta_2\). (c)~Transformed AoA: incident \((\theta_u,\varphi_u)\) on facet \(p\) is re-expressed as \((\theta_{u,p},\varphi_{u,p})\) via rotation \((\eta_p,\beta_p)\) relative to \(\mathbf{n}_1\).\label{pic1}}
    \vspace{-2mm}
\end{figure}

A tessellated multiplanar CCAA with \(P\) polygonal facets is shown in Fig.~\ref{pic1}(a). Spherical curvature is approximated by hexagonal tiling, with pentagonal facets added for closure. 
\vspace{-3mm}
Each facet hosts a co-planar CCAA aperture consisting of $V$ rings with $M_v$ elements at the $v$-th ring, i.e., $v\in \{1,\cdots,V\}$. The analysis extends to spherically curved assemblies provided the shielding condition holds. Without loss of generality, let facet $p=1$ be the reference with $\mathbf{n}_1$ aligned with the $y$-axis. Each facet normal $\mathbf{n}_p$ is a unit vector on $S^2$, where $S^2$ is the unit sphere in $\mathbb{R}^3$. 

In an uplink scenario with $U$ users, the angle of arrival (AoA) for user $a \in \{1, \dots, U\}$ is mapped to the reference facet $p=1$ through rotation $(\eta_p, \beta_p)$, resulting in the transformed AoA $(\theta_{a,p}, \varphi_{a,p})$ shown in Figs.~\ref{pic1}(b) and \ref{pic1}(c). The aggregate channel vector at the base station (BS) for user $a$ is 
\begin{equation}
    \hat{\mathbf{h}}_a = \sqrt{\xi_{a}}\mathbf{h}_a,
\end{equation} 
where $\xi_{a}$ is the deterministic large-scale channel attenuation (path loss and shadowing) between user $a$ and the BS, and
\vspace{-1mm}
\begin{equation}
  \mathbf{h}_a = \begin{bmatrix}\mathbf{h}_{a,1}^T & \mathbf{h}_{a,2}^T & \dots & \mathbf{h}_{a,P}^T \end{bmatrix}^T\;\in\;\mathbb{C}^{N_{\mathrm{tot}}\times 1}.
  \label{R1} \vspace{-1mm}
\end{equation}
In \eqref{R1}, $N_{\mathrm{tot}} = P M$ is the number of elements of the ST-CCAA with $M = \sum_{v=1}^V M_v$; $\mathbf{h}_{a,p} = g_{a,p} \bigl[
    e^{\jj \PSI{1}{1}{a}{p}},\dots,
    e^{\jj \PSI{M_v}{V}{a}{p}}
  \bigr]^T$ denotes the steering vector of the $p$-th CCAA facet as seen by user $a$; $g_{a,p}$ is a facet-visibility coefficient specified as $g_{a,p}{=}0$ for back-facing facets ($\mathbf{n}_p^{T}\hat{\mathbf{k}}_a \le 0$) and $g_{a,p}{=}1$ for forward-facing facets (ideal back-face isolation is assumed throughout); $\mathbf{k}_{a}=\tfrac{2\pi}{\lambda}\hat{\mathbf{k}}_a$ is user~$a$'s wave vector along the unit arrival direction $\hat{\mathbf{k}}_a$; $\PSI{m}{v}{a}{p}=
  2\pi R_v\sin\theta_{a,p}
  \cos\!\bigl(\varphi_{a,p}-\varphi_v^m\bigr)
  +\Phi_{p,\text{offset}}$ is the phase of element $m$ in ring $v$; $R_v=d_v/[2\sin(\pi/M_v)]$ is the ring radius (all lengths $d_v$, $d_z$, $R_v$, $\mathbf{r}$, the aperture diameter $D$, and the user range $r_u$ are in units of $\lambda$, i.e., $\lambda{=}1$ and $k_0{=}2\pi$; $\lambda$ is kept explicit in spacing conditions for readability), $\Phi_{p,\text{offset}}=\mathbf{k}_{a}^{T}\mathbf{r}_{c,p}$ is the facet-centre phase offset; and $\mathbf{r}_{c,p}$ is facet~$p$'s centre position.
The transformed AoA satisfies
\begin{equation}
\begin{split}
&\cos\theta_{a,p}={} \cos\theta_a\cos\eta_p + \sin\theta_a\sin\eta_p\cos\!\bigl(\varphi_a-\beta_p\bigr),\\
&\varphi_{a,p}={} \operatorname{atan2}\!\left( \begin{aligned} &\sin\theta_a\sin(\varphi_a-\beta_p),\\ &\sin\theta_a\cos\eta_p\cos(\varphi_a-\beta_p) -\cos\theta_a\sin\eta_p \end{aligned} \right).
\end{split}
\label{R3}\vspace{-3 mm}
\end{equation}
The received signal at BS is $\mathbf{y}=\sum_{a=1}^U \sigma_{xa}\hat{\mathbf{h}}_a x_a +\mathbf{z}$, where $x_a$ is the unit-power user $a$ symbol, $\sigma^2_{xa}$ its transmit power, and $\mathbf{z}\sim\mathcal{CN}(\mathbf{0},\sigma_0^2\mathbf{I}_{N_{\mathrm{tot}}})$ is AWGN, with $\mathbf{y},\hat{\mathbf{h}}_a,\mathbf{z}\in\mathbb{C}^{N_{\mathrm{tot}}\times1}$. \vspace{-4mm}
\section{FP Analysis of the P-Faceted ST-CCAA}
Unlike the Kronecker-separable UCLA baseline in~\cite{firoz}, the proposed ST-CCAA channel vector is non-separable owing to the facet-specific tilt $\eta_p$. From \eqref{R1}, the normalised inner product $\alpha_{a,b}$ between users $a,b \in \{1,\dots,U\}$ is \vspace{-2mm}

\begin{equation}
    \alpha_{a,b} = \frac{1}{N_{\mathrm{nor}}} \mathbf{h}_a^H \mathbf{h}_b
    = \frac{1}{N_{\mathrm{nor}}} \sum_{p=1}^{P}
      \mathbf{h}_{a,p}^H \mathbf{h}_{b,p},
    \label{R5}
\end{equation}
where $N_{\mathrm{nor}}=M\bigl({(\textstyle\sum_{p=1}^P g_{a,p})(\textstyle\sum_{p=1}^P g_{b,p})}\bigr)^{0.5}=\|\mathbf{h}_a\|\,\|\mathbf{h}_b\|$. By the Cauchy--Schwarz inequality, $\alpha_{a,a}=1$ and $|\alpha_{a,b}|\le 1$.
\vspace{-2mm}
\begin{remark}[Shielding: Two Distinct Roles]
\label{rem:shield}
Facet shielding is a modelling assumption underpinning the block form of~\eqref{R1}: back-facing facets ($\mathbf{n}_p^{T}\hat{\mathbf{k}}_a<0$) are taken to be isolated, while forward coupling between front facets is neglected under the far-field model of~\cite{firoz}. The relevant quantity is the antenna front-to-back and inter-facet isolation~\cite{balanis}, not electromagnetic-interference (EMI) shielding effectiveness; while the board- and material-level techniques of~\cite{Hybrid_BLS_Char, MXene_Shielding} indicate that adequate isolation is physically attainable, we treat back-face isolation as a modelling assumption and set $g_{a,p}\in\{0,1\}$. With binary visibility the active set is $\mathcal{A}_a\triangleq\{p:g_{a,p}=1\}=\{p:\mathbf{n}_p^T\hat{\mathbf{k}}_a>0\}$, of size $P_{\mathrm{active}}(\hat{\mathbf{k}}_a)\triangleq|\mathcal{A}_a|$; only these facets contribute to~\eqref{R1}, so $\|\mathbf{h}_a\|^2=M\sum_{p=1}^Pg_{a,p}=M\,P_{\mathrm{active}}(\hat{\mathbf{k}}_a)$ grows with the array, and two users couple only on the jointly active set $\mathcal{A}_{ab}\triangleq\mathcal{A}_a\cap\mathcal{A}_b$. A balanced, nearly antipodal tessellation gives $P_{\mathrm{active}}\le P/2$ ($\le P$ in general).
\end{remark}\vspace{-2mm}
Applying matched-filter beamforming $\mathbf{w}_a=\mathbf{h}_a/\|\mathbf{h}_a\|$ and treating other users as interference, the signal-to-interference-plus-noise ratio (SINR) for user $a$ is
\begin{equation}
    \mathrm{SINR}_a
    = \frac{\gamma_a}{\displaystyle\sum_{b=1,b\neq a}^{U}
      |\alpha_{a,b}|^{2}\,\gamma_b + 1},
    \label{R6}
\end{equation}
where $\gamma_a=\|\mathbf{h}_a\|^2 \xi_{a}\sigma^2_{xa}/\sigma^2_0$ is the single-user receive SNR of user $a$, $\forall a\in\{1,\cdots,U\}$. Similar to \cite{firoz}, the channel is normalised so that $\|\mathbf{h}_a\|^2=M\sum_{p=1}^P g_{a,p}$, i.e., the deterministic large-scale channel attenuation $\sqrt{\xi_{a}}$ is absorbed into the receive SNR $\gamma_a$.
We subsequently prove the FP of the ST-CCAA by showing that $\lim_{M\to\infty}|\alpha_{a,b}|=0$, $\forall b \neq a$.
\vspace{-3mm}
\subsection{FP Metric Decomposition and Convergence Proof}
Given the facet-user channel structure, the inner product on a jointly active facet $p\in\mathcal{A}_{ab}$ (where $g_{a,p}{=}g_{b,p}{=}1$ under ideal shielding; it vanishes otherwise) is
\begin{equation}
    \mathbf{h}_{a,p}^H \mathbf{h}_{b,p}
    = \sum_{v=1}^{V}\sum_{m=1}^{M_v}
       e^{\jj \left( \PSI{m}{v}{b}{p} - \PSI{m}{v}{a}{p} \right)}.
    \label{R7}
\end{equation}
Defining \(\Delta\mathbf{k}_{a,b}\triangleq\mathbf{k}_b-\mathbf{k}_a\), the phase difference is the projection of the wave-vector difference onto element position \(\mathbf{r}^{m,v}_p\):
\begin{equation}
  \Delta\psi^{m,v}_{p,(a,b)}
  = \psi^{m,v}_{b(p)} - \psi^{m,v}_{a(p)}
  = \Delta\mathbf{k}_{a,b}^T \mathbf{r}^{m,v}_{p}.
  \label{R8}
\end{equation}
Subtracting the centre offset $\mathbf{r}_{c,p}$ ensures that the element positions $\mathbf{r}^{m,v}_p$ reside within the facet plane. Let $\mathbf{R}(\eta_p,\beta_p) \in SO(3)$ be the rotation matrix mapping the local facet frame to the global frame; then, utilising the identity $\mathbf{u}^T \mathbf{R}\mathbf{v} = (\mathbf{R}^T\mathbf{u})^T \mathbf{v}$, the residual in-plane phase is derived as
\begin{equation}
  \delta_p
  = \frac{\bigl\|\Delta\mathbf{k}_{ab}^{(\text{in-plane},p)}\bigr\|}{k_0},
  \label{R9}
\end{equation}
where $k_0 = 2\pi/\lambda$ and $\Delta\mathbf{k}_{ab}^{(\text{in-plane},p)}$ is the projection of $\Delta\mathbf{k}_{ab}$ onto the plane of facet $p$. The terms $\delta_{1,p}$ and $\delta_{2,p}$ denote the local-frame Cartesian components of $\Delta\mathbf{k}_{a,b}/k_0$~\cite{firoz}, with $\delta_p = \mbox{$\sqrt{\delta_{1,p}^2 + \delta_{2,p}^2}$}$ representing the magnitude. Finally, defining the phase angle $\digamma_p = \arg(\delta_{2,p} + \jj\delta_{1,p})$ and the scaled radius $z_{v,p} = 2\pi R_v \delta_p$ yields the per-ring FP metric \vspace{-2mm}
\begin{equation}
  \bar{\alpha}_{v,p}
  = \frac{1}{M_v}\sum_{m=1}^{M_v}
    e^{\jj\, z_{v,p}\sin(\varphi_v^m+\digamma_p)}.
  \label{R10}
\end{equation}
\subsubsection{Analysis of Non-Separable Structure}
Since the ST-CCAA lacks Kronecker separability, the decomposition $\alpha_{a,b}=\alpha_{a,b}^v\alpha_{a,b}^h$ from~\cite{firoz} does not hold. Since the facet-centre offset $\Phi_{p,ab}\triangleq\Delta\mathbf{k}_{ab}^T\mathbf{r}_{c,p}$ is independent of $m$ and $v$, the position decomposition $\mathbf{r}^{m,v}_p = \mathbf{r}_{c,p} + \mathbf{R}(\eta_p,\beta_p)\tilde{\mathbf{r}}^{m,v}$, where $\tilde{\mathbf{r}}^{m,v}$ is the element coordinate in the local facet frame, allows~\eqref{R8} to be factored into a unit-modulus term $e^{\jj\Phi_{p,ab}}$ and a residual that reduces to $\bar{\alpha}_{v,p}$ per~\eqref{R9}--\eqref{R10}. This gives $\mathbf{h}_{a,p}^H \mathbf{h}_{b,p} = g_{a,p}^{*}g_{b,p}\,e^{\jj\Phi_{p,ab}}\sum_{v}M_v\bar{\alpha}_{v,p}$, nonzero only on $\mathcal{A}_{ab}$. With $\sum_{v=1}^V M_v=M$ and binary gains $g_{a,p}\in\{0,1\}$, the triangle inequality over rings gives $|\mathbf{h}_{a,p}^H\mathbf{h}_{b,p}|\le M\max_v|\bar{\alpha}_{v,p}|$ on $\mathcal{A}_{ab}$; summing over $\mathcal{A}_{ab}$, dividing by $N_{\mathrm{nor}}=\|\mathbf{h}_a\|\,\|\mathbf{h}_b\|=M\sqrt{P_{\mathrm{active}}(\hat{\mathbf{k}}_a)\,P_{\mathrm{active}}(\hat{\mathbf{k}}_b)}$, and using $|\mathcal{A}_{ab}|\le\sqrt{P_{\mathrm{active}}(\hat{\mathbf{k}}_a)\,P_{\mathrm{active}}(\hat{\mathbf{k}}_b)}$, one obtains
\vspace{-2mm}
\begin{equation}
  |\alpha_{a,b}|
  \le
  \max_{1\le p\le P,\,1\le v\le V}|\bar{\alpha}_{v,p}|.  
  \label{R12} \vspace{-0.5mm}
\end{equation}
As each $|\bar{\alpha}_{v,p}|$ corresponds to the interference metric of a planar UCA with fixed chord spacing $d_v>0$, the Bessel bounding technique is applicable per ring.

\textit{Probabilistic Motivation:} 
The tilt diversity renders the per-element phases in~\eqref{R7} approximately independent, so the sum behaves as a complex random walk with mean-square value \vspace{-2mm}
\begin{equation}
\mathbb{E}[|\alpha_{a,b}|^2] \propto \frac{1}{M},
\label{R13}\vspace{-1 mm}
\end{equation}
suggesting convergence in probability. As this does not establish pointwise convergence, we derive the deterministic bound.
\subsubsection{Per-Ring Convergence}
Following~\cite[Proposition~3, Case~2]{firoz}, we adopt \(d_v<\lambda/2\). Each \(\bar{\alpha}_{v,p}\) in~\eqref{R10} has the same form as the single-ring metric of~\cite{firoz}, with \(R_v\propto M_v\) under fixed \(d_v\). Substituting~\eqref{R10} into~\eqref{R12} with \(P,V\) fixed, for finite \(U\) with bounded \(\gamma_a\):
\vspace{-1mm}
\begin{theorem}[FP for ST-CCAA: Pointwise Convergence]
\label{thm:fp}
Consider an ST-CCAA with $P \ge 3$ facets whose outward normals $\{\mathbf{n}_p\}$ span $\mathbb{R}^3$, fixed chord spacings $0 < d_v < \lambda/2$, and $M_v \ge 2$ elements per ring. Assuming all user pairs satisfy $\delta_p > 0$ for $p\in \{1, \dots, P\}$, and given fixed inter-element spacing, negligible inter-facet coupling, and non-degenerate user directions, then as the array grows with $M_v\to\infty$ for every ring (with $M_{\min}$ growing in proportion to $M$), $|\alpha_{a,b}|\to0$ pointwise with rate $\mathcal{O}(M^{-1/3})$. Specifically, the explicit finite-$M$ bound is \vspace{-2mm}
\begin{equation}
\begin{split}
  |\alpha_{a,b}|
  \le
  \frac{2\nu_1-1}{\bigl(d_{\min}\,\delta_{\min}(P)\,M_{\min}\bigr)^{1/3}}
  + \frac{2(1/2)^{\nu_1 M_{\min}}}{1-(1/2)^{M_{\min}}},
\end{split}
  \label{R19}
\end{equation}
where $\nu_1=\lceil e\pi d_{\max}\rceil$ is independent of $M_v$, $M_{\min}=\min_v M_v$, and $\delta_{\min}(P)=\min_p\{\delta_p>0\}$ depends on the facet density and normal distribution. To ensure $|\alpha_{a,b}|\le\varepsilon$ for a threshold $\varepsilon>0$, inverting the $M_{\min}^{-1/3}$ term of \eqref{R19} shows it suffices that $M_{\min}\gg(2\nu_1-1)^3/(d_{\min}\delta_{\min}\varepsilon^3)$.
\end{theorem}
\begin{proof}
Apply the Bessel generating function $e^{\jj z\sin\phi}=\sum_{\iota=-\infty}^{\infty}J_\iota(z) e^{\jj\iota\phi}$ to~\eqref{R10}. With uniformly spaced elements at $\varphi_v^m=2\pi m/M_v$, exchanging summation order gives $\bar{\alpha}_{v,p}=\sum_{\iota}J_\iota(z_{v,p}) e^{\jj\iota\digamma_p}\cdot\frac{1}{M_v}\sum_{m=1}^{M_v} e^{\jj 2\pi m\iota/M_v}$. By discrete orthogonality, the inner sum is unity if $M_v$ divides $\iota$ and zero otherwise, so only integer multiples of $M_v$ survive. Separating the zero-order term and pairing positive and negative multiples via the identity $J_{-\iota}=(-1)^\iota J_\iota$, gives \vspace{-2mm}
\begin{equation}
\begin{split}
  \bar{\alpha}_{v,p}
  = &J_0(z_{v,p}) + \\
  &\sum_{\tau=1}^{\infty}
    \Bigl(e^{\jj\tau M_v\digamma_p}+(-1)^{\tau M_v}
    e^{-\jj\tau M_v\digamma_p}\Bigr)
    J_{\tau M_v}(z_{v,p}),
\end{split}
  \label{R14}
\end{equation}
for $\tau=1,2,\dots$. We bound $|\bar{\alpha}_{v,p}|$ by partitioning the sum at $\nu_1=\lceil e\pi d_{\max}\rceil$, where $d_{\max}=\max_v d_v$ and $d_{\min}=\min_v d_v$ are the maximum and minimum chord spacings:
\begin{equation}
  |\bar{\alpha}_{v,p}|\le \underbrace{|J_0(z_{v,p})|+2\sum_{\tau=1}^{\nu_1-1}|J_{\tau M_v}(z_{v,p})|}_{A_{1}} +\underbrace{2\sum_{\tau=\nu_1}^{\infty}|J_{\tau M_v}(z_{v,p})|}_{A_{2}}.
  \label{R15}
\end{equation}
For $A_1$, we employ the Landau bound $|J_n(x)|\le|x|^{-1/3}$~\cite{Landau2000}. Given $z_{v,p}=2\pi R_v\delta_p\ge d_{\min}\delta_{\min}M_v$ (since $\sin(\pi/M_v)\le\pi/M_v$), the $2\nu_1-1$ retained terms give
\begin{equation}
  A_1 \le \frac{2\nu_1-1}{\bigl(d_{\min}\delta_{\min}M_v\bigr)^{1/3}}.
  \label{R16}
\end{equation}
For $A_2$, we use Kapteyn's inequality $|J_n(nx)|\le|xe/2|^n$ for $|x|\le 1$~\cite{Watson1966}. With $\nu_1=\lceil e\pi d_{\max}\rceil$, the ratio $q=e\pi d_{\max}/2\nu_1\le 1/2$, yielding
\begin{equation}
  A_2 \le \frac{2(1/2)^{\nu_1 M_v}}{1-(1/2)^{M_v}}
  \xrightarrow{M_v\to\infty} 0.
  \label{R17}
\end{equation}
Combining \eqref{R16} and \eqref{R17}, we find that for each $\delta_p>0$,
\vspace{-2mm}
\begin{equation}
  |\bar{\alpha}_{v,p}| \le \frac{2\nu_1-1}{\bigl(d_{\min}\,\delta_{\min}\,M_v\bigr)^{1/3}} + \frac{2(1/2)^{\nu_1 M_v}}{1-(1/2)^{M_v}}\xrightarrow{M_v\to\infty} 0,
  \label{R18}
\end{equation}
implying $|\bar{\alpha}_{v,p}|=\mathcal{O}(M_v^{-1/3})$ as $M_v\to\infty$. Substituting \eqref{R18} into \eqref{R12}, with the worst ring $M_{\min}$, yields \eqref{R19}. \vspace{-2mm}
\end{proof}
\vspace{-2mm}
\begin{corollary}[SINR Favourable Propagation]
\label{cor:sinr}
Under the conditions of Theorem~\ref{thm:fp}, with fixed per-user receive SNRs $\gamma_u$, for any finite number of users $U$ it follows that $\lim_{M\to\infty}\mathrm{SINR}_a=\gamma_a$.\vspace{-2mm}
\end{corollary}
\begin{proof}
From Theorem~\ref{thm:fp}, $|\alpha_{a,u}|\to 0$ for all $u \neq a$, which implies that the aggregate interference $\sum_{u=1,u \neq a}^{U}|\alpha_{a,u}|^2\gamma_u \to 0$. Substituting this result into \eqref{R6} yields $\mathrm{SINR}_a\to\gamma_a$. 
\end{proof}
\vspace{-3mm}
\begin{remark}[Scope and Validity]
\label{rem:scope}
For fixed $\hat{\mathbf{k}}_a$, each facet gives at most one
distinct partner with $\delta_p=0$, namely
$\hat{\mathbf{k}}_b=\hat{\mathbf{k}}_a
-2(\mathbf{n}_p^T\hat{\mathbf{k}}_a)\mathbf{n}_p$.
The resulting finite union of reflection graphs has measure zero in $S^2\times S^2$; hence, Theorem~\ref{thm:fp} holds for almost all user configurations. The proof requires $\delta_p>0$ only on the finite jointly active set, while sphericity supplies generic normal diversity and wide-angle coverage. Common-boresight directional elements induce bounded per-facet weights, changing only the constant in~\eqref{R19}, not the $\mathcal{O}(M^{-1/3})$ rate. The bound is non-uniform as
$\delta_{\min}\to0$ and does not cover imperfect isolation or appreciable inter-facet coupling, fixed-aperture scaling, or highly irregular tessellations. At fixed spacing $d_v$, the aperture diameter $D$ scales linearly with $M$~\cite{firoz}, so the Fraunhofer distance $d_{\mathrm{F}}=2D^2 \lambda$ scales as $\Theta(M^2)$. For practical deployments with a fixed user range $r_u$, the explicit non-asymptotic bound in~\eqref{R19} remains strictly valid for every finite $M$ for which the user remains in the far field, $d_{\mathrm{F}}\le r_u$. Asymptotically, the far-field assumption, fixed spacing, and the $M\to\infty$ limit remain mutually consistent under the scaling condition $r_u=\Omega(M^2)$. In the radiating near field, the additional range-dependent Fresnel phase term provides depth-domain focusing that aids user separation rather than destroying the cancellation~\cite{cui,bjorn}. Full spherical-wavefront near-field analysis is deferred to future work.\vspace{-5mm} 
\end{remark}
\subsection{Spectral Characterisation and Grating-Lobe Comparison}
\label{sec:spectral}
In the inter-user interference (IUI) context, a grating lobe occurs if $|\alpha_{a,b}|$ converges to a non-zero constant as $M\to\infty$, violating FP. Per Theorem~\ref{thm:fp}, this happens when $\delta_p=0$, i.e., $\Delta\mathbf{k}_{a,b}$ is parallel to $\mathbf{n}_p$. The two-level analysis below gives a spectral interpretation complementary to Theorem~\ref{thm:fp}; it is a qualitative diagnostic and is not used in the convergence proof.
\subsubsection{Derivation of the Structure Factor $\mathcal{S}_{lq}$}
\label{sec:slq}
Applying the Jacobi--Anger expansion \cite{Watson1966},
\begin{equation}
  e^{\jj\mathbf{k}^T\mathbf{r}}
  =4\pi\!\sum_{l=0}^{\infty}\sum_{q=-l}^{l}
  \jj^l\sphbess_l(kr)[\Ylm(\hat{\mathbf{k}})]^*\Ylm(\hat{\mathbf{r}}),
  \label{R20}
\end{equation}
where $k=\lvert\mathbf{k}\rvert$, $r=\lvert\mathbf{r}\rvert$, $\hat{\mathbf{k}} = \frac{\mathbf{k}}{\lvert \mathbf{k} \rvert}$,  and $\hat{\mathbf{r}} = \frac{\mathbf{r}}{\lvert \mathbf{r} \rvert}$ to the phase differences in~\eqref{R7}--\eqref{R8}, the mode content of the FP metric is captured by the diagnostic expansion: \vspace{-2mm}
\begin{equation}
\begin{split}
  \alpha_{a,b}
  \approx\frac{4\pi}{\sqrt{P_{\mathrm{active}}(\hat{\mathbf{k}}_a)\,P_{\mathrm{active}}(\hat{\mathbf{k}}_b)}}\!\sum_{l,q}
  \jj^l[\Ylm(\Omega_{a,b})]^*
  \mathcal{S}_{lq},
\end{split}
\label{R21}
\vspace{-4mm}
\end{equation}
where the jointly active structure factor is $\mathcal{S}_{lq} \triangleq \frac{1}{M}\sum_{p\in\mathcal{A}_{ab}}\sum_{v=1}^{V} \sum_{m=1}^{M_v} \sphbess_l(\kappa_v) \Ylm(\Omega_{p,m})$, with $\kappa_v \triangleq |\Delta\mathbf{k}_{a,b}|R_v$, $\Omega_{a,b} \triangleq \Delta\mathbf{k}_{a,b}/|\Delta\mathbf{k}_{a,b}|$, and $\Omega_{p,m} \triangleq (\theta_{p,m},\varphi_{p,m})$. The tilt $(\eta_p,\beta_p)$ is implicitly contained in $\Ylm(\Omega_{p,m})$. As a diagnostic, the prefactor is immaterial; only the mode content of $\mathcal{S}_{lq}$ governs grating behaviour.
\subsubsection{Level 1: Within-Facet Planar Array Factor}
\label{sec:level1}
Each facet $p$ is a planar CCAA sub-array tilted at $(\eta_p,\beta_p)$. Since the concentric-ring geometry is Bessel-class, it requires no Nyquist condition to avoid aliasing; however, a local planar approximation lets us use the PSF as a diagnostic (a CCAA is not itself a Bravais lattice; the lattice is only a local sampling analogy). For a 2D Bravais lattice $\mathcal{L}$ \cite{bened}, the density is \vspace{-2mm}
\begin{equation}
  \sum_{\mathbf{r}_n\in\mathcal{L}}\delta(\mathbf{r}-\mathbf{r}_n)
  =\frac{1}{\mathrm{vol}(\mathcal{L})}
  \sum_{\mathbf{G}\in\mathcal{L}^{\star}}e^{-\jj\mathbf{G}\mathbf{r}},
  \label{R22}
\end{equation}
where $\mathcal{L}^{\star}$ is the reciprocal lattice. The reciprocal primitive vectors $\mathbf{b}_1,\mathbf{b}_2$ satisfy $\mathbf{a}_\mu^T\mathbf{b}_\chi = 2\pi\delta_{\mu\chi}$ for $\mu,\chi\in\{1,2\}$:
\begin{equation}
  \mathbf{b}_1=\frac{2\pi}{\mathrm{vol}(\mathcal{L})}
  (\mathbf{a}_2\times\hat{z}),\quad
  \mathbf{b}_2=\frac{2\pi}{\mathrm{vol}(\mathcal{L})}
  (\hat{z}\times\mathbf{a}_1),
  \label{R23}
\end{equation}
where $\hat{z}$ is the unit normal to the lattice plane. The resulting within-facet array factor is
\begin{equation}
  AF_p(\boldsymbol{\digamma})
  =AF_{\mathrm{cont}}(\boldsymbol{\digamma})
  +\sum_{\mathbf{G}\in\mathcal{L}^{\star},\,\mathbf{G}\ne0}
   AF_{\mathrm{cont}}(\boldsymbol{\digamma}-\mathbf{G}),
  \label{R24}
\end{equation}
where $\mathbf{G}$ denotes the displacement of aliasing replicas. For $d_v < \lambda/2$, $|\mathbf{G}|_{\min} > 2k_0$, pushing replicas outside the visible region. Here $|\mathbf{G}|_{\min}$ is the smallest nonzero reciprocal-lattice vector magnitude, so the strict inequality $d_v<\lambda/2$ keeps every replica outside the visible region.

\subsubsection{Level 2: Inter-Facet Array Factor on $S^2$}
\label{sec:level2}
Because facet-centre directions $\{\Omega_{c,p}\}_{p=1}^P$ on $S^2$ are non-periodic, the PSF is inapplicable. Instead, inter-facet contributions to $\mathcal{S}_{lq}$ are governed by the mode-$(l,q)$ content of the facet distribution:
\vspace{-2mm}
\begin{equation}
  \widetilde{Y}_{lq}
  \triangleq
  \sum_{p=1}^{P}\Ylm(\Omega_{c,p}).
  \label{R25}
  \vspace{-2mm}
\end{equation}
As $P$ increases, the characteristic separation $\Delta\Omega_c \approx \sqrt{4\pi/P}$ decreases, suppressing higher-order modes $l \gtrsim \sqrt{\pi P}/2$. Modes with $l > k_0 R_{\mathrm{ap}}$ are rapidly decaying under the far-field model in~\eqref{R1}, where $R_{\mathrm{ap}}$ is the outer aperture radius. Thus, provided within-facet aliasing is controlled via Theorem~\ref{thm:fp}, $\widetilde{Y}_{lq}$ serves as a diagnostic for residual inter-facet aliasing that contributes negligibly to the array factor. Complete suppression of propagating modes requires $\widetilde{Y}_{lq}=0$ for $1\le l\le k_0 R_{\mathrm{ap}}$, which requires a spherical $l$-design at each degree $l$; near-uniform (e.g., geodesic) tessellations approach this.
\begin{remark}[Design Trade-off: Facet Density vs. Convergence Rate]
\label{rem:tradeoff}
The finite-$M$ bound in~\eqref{R19} is governed by the minimum normalised in-plane separation $\delta_{\min}(P)$. As facet normals $\{\mathbf{n}_p\}$ densify on $S^2$, the probability of $\mathbf{n}_{p^*}$ aligning with any $\Delta\mathbf{k}_{a,b}$ increases, driving $\delta_{\min}(P)\to 0$ and increasing the required $M_{\min}$. Thus, while larger $P$ improves angular diversity and spectral smoothing, it tightens the per-ring convergence requirement.
\end{remark}\vspace{-2mm}
\subsubsection{Grating-Lobe Comparison}
We compare the grating-lobe performance of the UCLA and ST-CCAA architectures. 

\textbf{UCLA:} Vertical translational symmetry implies $AF_{\mathrm{UCLA}}=AF_{\mathrm{CCAA}}\cdot AF_{\mathrm{ULA}}$ ~\cite[Proposition~2]{firoz}, meaning all $N_c$ rings share the same transformed AoA and lack inter-ring diversity. Here the 1D PSF on $\mathcal{L}_z= \Lambda d_z$ ($\forall \Lambda \in \mathbb{Z}$) is exact, since the vertical stack is a uniform 1-D lattice, giving an exact array-factor description and the closed-form grating-lobe condition \vspace{-2mm}
\begin{equation}
  \cos\theta_{GL}=\cos\theta_0+\frac{\zeta\lambda}{d_z},
  \quad \zeta\in\mathbb{Z}\setminus\{0\},
  \label{R28}
\end{equation}
which enters the visible region for $d_z\ge\lambda$. Since $AF_{\mathrm{CCAA}}\ne 0$ across all $\varphi$ at these elevations, each replica manifests as a persistent full-azimuth ridge that is sharpened, but not displaced, as $N_c$ increases~\cite{firoz}.

\textbf{ST-CCAA:} Conversely, each facet $p$ presents distinct transformed AoAs, yielding facet-specific $\delta_p$ via~\eqref{R9}. Using PSF as a structural diagnostic, a global IUI grating lobe requires \vspace{-2mm}
\begin{equation}
  \frac{2\pi m_p}{d_v}
  =\Delta\mathbf{k}_{a,b}^T\hat{\mathbf{e}}_{v,p},
  \quad m_p\in\mathbb{Z}\setminus\{0\},\;\forall\,p,
  \label{R29}
\end{equation}
where $\hat{\mathbf{e}}_{v,p}$ is the in-plane radial direction on facet $p$. Since $d_v<\lambda/2$ precludes the single-facet case and $\hat{\mathbf{e}}_{v,p}$ varies across the sphere, \eqref{R29} admits no consistent solution. For almost all directions, at most one facet aligns with $\Delta\mathbf{k}_{a,b}$, while others decay per Theorem~\ref{thm:fp}. Under binary visibility an aligned facet ($\delta_{p_0}=0$) is front-facing for one user and back-facing for the other (the projections $\mathbf{n}_{p_0}^T\hat{\mathbf{k}}_a$ and $\mathbf{n}_{p_0}^T\hat{\mathbf{k}}_b$ have opposite signs), hence not jointly active; every jointly active facet has $\delta_p>0$. Writing $A_0(M):=\tfrac{1}{M}|\sum_{v=1}^{V} M_v \bar{\alpha}_{v,p_0}|$ for the jointly active facet $p_0$ of least separation, the grating response over a common active set of size $P_{\mathrm{active}}$ is bounded by
\begin{equation}
  |\alpha_{a,b}^{GL}(M)|
  \le
  \frac{A_0(M)+(P_{\mathrm{active}}-1)C^\star M^{-1/3}}
       {P_{\mathrm{active}}},
  \label{R30}
\end{equation}
where $C^\star=(2\nu_1-1)/(d_{\min}\delta_{\min})^{1/3}$. Since each jointly active facet satisfies $|\bar{\alpha}_{v,p_0}| \le C^\star M^{-1/3}$
and $\sum_{v=1}^{V} M_v = M$, we obtain
$A_0(M)
= \frac{1}{M}\Bigl|\sum_{v=1}^{V} M_v \bar{\alpha}_{v,p_0}\Bigr|
\le C^\star M^{-1/3},$
i.e., \ $A_0(M)=\mathcal{O}(M^{-1/3})$. Hence, by~\eqref{R30},
$|\alpha_{a,b}^{\mathrm{GL}}(M)| = \mathcal{O}(M^{-1/3})$ and the bound vanishes as
$M\to\infty$.

\begin{remark}[Source of Grating Suppression]
\label{rem:source}
The ST-CCAA and UCLA are locally identical (sharing the same CCAA ring geometry and $d_v < \lambda/2$). Grating suppression in ST-CCAA is therefore a global consequence of normal diversity on $S^2$, beyond the local sub-half-wavelength condition it shares with the UCLA.
\end{remark}\vspace{-2mm}
\textit{Design Duality: }
The contrast between the persistent unit-amplitude ridge of the UCLA~\eqref{R28} and the inconsistent, decaying residual of the ST-CCAA~\eqref{R30} establishes the following: \vspace{-2mm}
\begin{equation}
  \underbrace{d_z<\lambda}_{\text{UCLA: no ridge}}
  \;\;\text{vs.}\;\;
  \underbrace{\text{aperiodic facet normals on }S^2}
             _{\text{ST-CCAA: reduced grating content}},
  \label{R31}
\end{equation}
where the former relies on frequency-domain exclusion ($2\pi/d_z>2k_0$) and the latter on the absence of a periodic stacking direction on $S^2$.
\vspace{-3mm}
\subsection{Comparison with the UCLA Baseline}
Table~\ref{Tab1} contrasts the two arrays under \(d_v<\lambda/2\), \(M\to\infty\).
\vspace{-6mm}
\begin{table}[h]
\caption{ST-CCAA vs. UCLA~\cite{firoz} Comparison}
\vspace{-2mm}
\label{Tab1}
\centering
\scriptsize
\setlength{\tabcolsep}{2pt}
\begin{tabular}{|p{15mm}|p{30mm}|p{39mm}|}
\hline
\textbf{Feature}
  & \textbf{UCLA~\cite{firoz}}
  & \textbf{ST-CCAA}\\
\hline
Structure
  & Separable; parallel rings; identical normals
  & Non-separable; distinct tilts \((\eta_p,\beta_p)\); diverse normals on \(S^2\)\\
\hline
FP mechanism
  & Kronecker factorisation
  & Triangle inequality~\eqref{R12} per facet\\
\hline
Bounding
  & Bessel, Landau, Kapteyn~\cite{firoz,Landau2000,Watson1966}
  & Same per ring\\
\hline
Spacing
  & \(d_v<\lambda/2\) (Prop.~3 Case~2~\cite{firoz})
  & Same\\
\hline
FP rate
  & \(\mathcal{O}(N_c^{-1/3})\) pointwise
  & \(\mathcal{O}(M^{-1/3})\) (Thm.~\ref{thm:fp})\\
\hline
SINR
  & \(\to\gamma_1\) (Prop.~3~\cite{firoz})
  & \(\to\gamma_1\) (Cor.~\ref{cor:sinr})\\
\hline
Grating lobe
  & 1D PSF exact on \(\mathcal{L}_z\); coherent ridge~\eqref{R28};
    full amplitude; fixed by \(d_z\)
  & PSF inapplicable to \(S^2\); generically inconsistent~\eqref{R29};
    bounded residual~\eqref{R30}, decays \(\mathcal{O}(M^{-1/3})\)\\
\hline
Grating source
  & Periodic stacking; all rings coherent
  & Normal diversity \(\mathbf{n}_p\) on \(S^2\); facets incoherent\\
\hline
FP failure
  & \(\phi_a=\phi_b\), \(\theta_a=\theta_b\)
  & \(\Delta\mathbf{k}_{a,b}\) parallel to \(\mathbf{n}_p\) (measure-zero)\\
\hline
\end{tabular}
\vspace{-5mm}
\end{table}

\subsection{Numerical Results}
\label{sec:numerical}
\begin{figure*}[t]
\centering
\includegraphics[width=\textwidth,height=65mm,keepaspectratio]{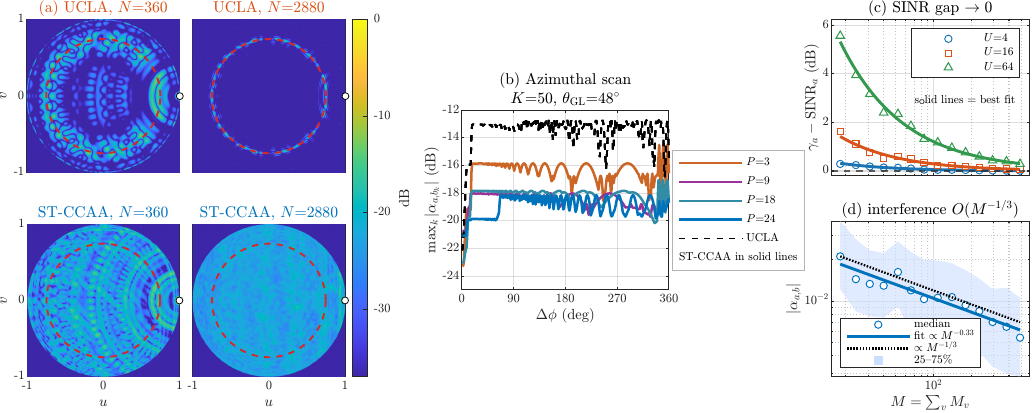}
\vspace{-3mm}
\caption{ST-CCAA vs.\ UCLA at equal element count $N{\equiv}N_{\mathrm{tot}}$ ($=PM$ facets\,/\,$N_cM$ levels); $M{=}\sum_v M_v{=}120$ ($M_v{=}40$, $V{=}3$, $d_v{=}0.45\lambda$, $d_z{=}1.50\lambda$) is fixed in (a)--(b), so $N{=}360$ and $2880$
correspond to $P{=}N_c{=}3$ and $24$. \emph{Grating lobes}: reference user at $(90^\circ,0)$ (white marker), giving $\theta_{\mathrm{GL}}{\approx}48^\circ$ by~\eqref{R28} (red dashed circle). (a)~$20\log_{10}|\alpha_{a,b}|$ in $uv$-space (top UCLA, bottom ST-CCAA); (b)~worst-case $|\alpha_{a,b}|$ over $K{=}50$ users swept in azimuth $\Delta\phi$ at $\theta_{\mathrm{GL}}$ (ST-CCAA solid, UCLA black
dashed, $N_c$-invariant). \emph{Favourable propagation}, (c)--(d): $P{=}20$, $V{=}3$, $\gamma_a{=}10$~dB, isotropic users, shared abscissa
$M$; markers simulated, lines fitted. (c)~SINR gap $\gamma_a-\mathrm{SINR}_a$ for $U\in\{4,16,64\}$; (d)~median $|\alpha_{a,b}|$ with 25--75\% band; fitted slope $-0.33$ vs.\ $M^{-1/3}$.}

\label{fig:results}
\vspace{-6mm}
\end{figure*}
Fig.~\ref{fig:results} corroborates the analysis using~\eqref{R5}; settings in the caption. In Fig.~\ref{fig:results}(a) the UCLA ring at $\sin\theta_{\mathrm{GL}}$ predicted by~\eqref{R28} persists at $N{=}360$ and $N{=}2880$, whereas the ST-CCAA shows none, the facet-wise variation of $\hat{\mathbf{e}}_{v,p}$ rendering~\eqref{R29} inconsistent. In Fig.~\ref{fig:results}(b) the UCLA worst case stays near $-13$~dB while the ST-CCAA falls once $P\ge9$, following the $A_0(M)/P_{\mathrm{active}}$ term of~\eqref{R30}: suppression is a facet-normal-diversity effect at fixed $M$. Figs.~\ref{fig:results}(c)--(d) validate Corollary~\ref{cor:sinr} and Theorem~\ref{thm:fp}: the gap $\gamma_a-\mathrm{SINR}_a$ vanishes for every $U$, and the median $|\alpha_{a,b}|$ decays with slope $\approx-0.33$, matching the $M^{-1/3}$ rate of~\eqref{R18} across the 25--75\% band.
\vspace{-5mm}
\section{Conclusion}
This letter establishes pointwise FP for the ST-CCAA by replacing Kronecker factorisation with a per-facet triangle inequality, reducing the non-separable spherical geometry to UCA sub-problems in which within-facet aliasing is controlled by Bessel decay. Spectral analysis via the structure factor $\mathcal{S}_{lq}$ and the PSF heuristically explains the absence of coherent inter-facet grating ridges. The grating residual stays strictly below unity and decays as $\mathcal{O}(M^{-1/3})$, due to the 3-D facet-normal diversity. Despite trade-offs between peak gain and hemispherical coverage, the ST-CCAA is well suited for wide-angle non-terrestrial networks (NTN) and terrestrial deployments. Future work will refine the $\mathcal{O}(M^{-1/3})$ bound
via phase diversity and extend the framework to near-field XL-MIMO.
 \vspace{-3mm}
\bibliographystyle{IEEEtran}
\bibliography{Bibliography}
\end{document}